\documentclass{article}
\usepackage[utf8]{inputenc}
\usepackage{textcomp}
\usepackage{amsmath}
\usepackage{amssymb}
\usepackage{graphicx}
\usepackage{braket}
\usepackage{url} 
\usepackage{comment} 
\usepackage{subfigure}
\usepackage[makeroom]{cancel}

\usepackage[font=footnotesize]{caption}  

  \usepackage[
    backend=biber,
    style=authoryear,
  ]{biblatex}
\usepackage{authblk} 

\begin{document}

\title{\huge \vspace{-2.0cm} Nonlocal action in Everettian Quantum Mechanics \\[0.5em]
       \large For the volume:\\ \textit{Local Quantum Mechanics: Everett, Many Worlds, and Reality} \\ (Oxford University Press, forthcoming)}

\author[1]{Mordecai Waegell}
\author[1,2]{Kelvin J. McQueen}

\affil[1]{\small Institute for Quantum Studies, Chapman University}
\affil[2]{\small Philosophy Department, Chapman University}


\date{\today}

\maketitle

   \begin{abstract}
According to a common view, Everettian quantum mechanics (EQM) is a local theory because it avoids nonlocal action at a distance, and this is an important point in EQM's favor. Unlike collapse theories, EQM does not allow an action on one system to change the reduced density matrix (RDM) of a remote entangled system—a clear case of nonlocal action. However, EQM does allow an action on one system to change the global state of the system and its remote entangled partners. We argue that such changes should also count as nonlocal actions, meaning EQM is not local after all. First, we consider an argument to the contrary, which deems such global changes to be mere \textit{extrinsic} changes, whereas nonlocal action requires \textit{intrinsic} changes to the remote system. We respond that the intrinsic-extrinsic distinction is problematic and cannot hold the weight of this argument. We then try to clarify when actions that change global states count as nonlocal actions. We argue that it is when the global states are essential explanatory mechanisms of the theory. In EQM, the global state is needed to explain why, in an anti-correlated Bell state, Alice’s measuring spin-up ensures that she encounters only the branch where Bob measures spin-down. 
      \end{abstract}

\newpage



\section{Introduction}

According to a common view, Everettian quantum mechanics (EQM) is a local theory because it avoids nonlocal action at a distance.\footnote{\textcite{Timpson2002, Bacciagaluppi2002, deutsch2012vindication, wallace2012emergent, tipler2014quantum, Vaidman2016, Brown2016, ney2023argument, faglia2024non}, and several chapters in this volume.} This is often taken to count in EQM's favor, with locality being a desirable feature of a physical theory. In this paper, we raise doubts about this view, and argue that while EQM avoids the nonlocal action indicative of wavefunction collapse, it still seems committed to other kinds of nonlocal action, which affect certain global states. 

For example, in an anti-correlated Bell state, Alice measuring her particle is an action that nonlocally changes the global state that includes Bob's remote particle.  The global state changes in a way that ensures her successor who saw spin-up will be in a branch where, if Bob measured along the same axis, his successor saw spin-down. We think this is nonlocal because the altered global state plays an essential explanatory role in EQM, in the sense that it is used to determine the outcome seen by each Bob, and thus to assure the proper entanglement correlations with the Alices. Therefore, we will argue that while EQM avoids one specific type of nonlocal action, it does not avoid nonlocal action in general. 

In collapse theories, a local measurement on one part of an entangled system can instantaneously change the reduced density matrix (RDM) of the remote system. To illustrate, consider a pair of qubits prepared in the Bell state:
\begin{equation}
    \ket{\Psi} = \frac{1}{\sqrt{2}} (\ket{0}_A \ket{1}_B - \ket{1}_A \ket{0}_B).
\end{equation}
Suppose an observer, Alice, measures qubit \( A \) in the computational basis \( \{ \ket{0}, \ket{1} \} \). In a collapse theory, upon measurement, the wavefunction collapses to either \( \ket{0}_A \ket{1}_B \) or \( \ket{1}_A \ket{0}_B \) with equal probability. Consequently, the RDM of qubit \( B \), obtained by tracing out system \( A \), transitions from a maximally mixed state:
\begin{equation}
    \rho_B = \frac{1}{2} \ket{0} \bra{0} + \frac{1}{2} \ket{1} \bra{1}
\end{equation}
to a pure state, either \( \ket{0} \bra{0} \) or \( \ket{1} \bra{1} \), depending on Alice’s outcome. This is an instance of nonlocal action: Alice’s measurement performed in one region instantaneously alters the remote system’s RDM.

By contrast, in EQM, wavefunction collapse does not occur. Instead, the measurement leads to branching, where both outcomes exist in superposition, each correlated with a different copy of the observer:
\begin{equation}
    \ket{\Psi'} = \frac{1}{\sqrt{2}} (\ket{0}_A \ket{1}_B \ket{\text{Alice}_0} - \ket{1}_A \ket{0}_B \ket{\text{Alice}_1}).
\end{equation}
Here, Alice herself undergoes a branching process, with one copy observing \( 0 \) and another observing \( 1 \). Since both branches coexist and there is no collapse, the reduced density matrix of qubit \( B \) remains unchanged at (2). Thus, EQM avoids the nonlocal action at a distance involved in wavefunction collapse, because no single outcome is realized globally—rather, all outcomes exist in a superposition of branches, and the RDM remains invariant.

The above contrast may make it seem that EQM is a local theory that avoids nonlocal action at a distance. But all that has been shown is that EQM avoids one type of nonlocal action, one that involves changes in remote RDMs. But what is so special about RDMs? EQM may still entail nonlocal actions that make real changes in something else. It seems to us that EQM allows other forms of nonlocal actions. In particular, ones that change certain \textit{global states} of the measured system and its remote entangled partners. 

An example is the transition from (1) to (3), where as a result of Alice's action, qubit $B$'s quantum state is described by a global state that includes Alice. This change in $B$'s global state does not change $B$'s RDM. Since $B$'s RDM encodes the probabilities of the outcomes of measurements of local observables of $B$, such measurements cannot reveal any change in $B$. But we think that $B$ has still undergone a real change. For this change makes a difference when Alice meets Bob. It ensures that whenever Bob (after measuring $B$) is ready to meet Alice, Bob$_0$ will meet Alice$_1$ and Bob$_1$ will meet Alice$_0$. To illustrate, imagine that Bob, unaware of Alice’s measurement, measures his qubit in the same computational basis. This results in:

\begin{equation}
    \ket{\Psi''} = \frac{1}{\sqrt{2}} (\ket{0}_A \ket{1}_B \ket{\text{Alice}_0} \ket{\text{Bob}_1} - \ket{1}_A \ket{0}_B \ket{\text{Alice}_1} \ket{\text{Bob}_0}).
\end{equation}

The reduced density matrices for Alice and Bob's respective qubits remain invariant, ensuring that local measurement statistics remain the same. However, the global state has changed in a way that ensures a perfect correlation when Alice and Bob meet. 

On the face of it, this is a form of nonlocal action: Alice’s local measurement has restructured the global wavefunction in such a way that Bob’s future interaction with Alice is guaranteed to align with the correct branch.

In the next section (section \ref{argument}), we consider an influential argument to the contrary, which aims to show that there is no nonlocal action in EQM. A key premise of the argument is that Alice's action makes no \textit{real change} to $B$, because it does not change any \textit{intrinsic} properties of $B$. The idea is that only changes to intrinsic properties matter when it comes to nonlocal actions, and while RDMs are intrinsic properties (of remote entangled subsystems), global states are not. 

In response, we note that these arguments lack a clear definition for the central notion of `intrinsic'. We show that the intuitive notion does not apply in a clear way even to classical metaphysics, let alone quantum mechanics, making its use in categorizing physical theories as local or not questionable. Ultimately, we concur with \textcite{marshall_weatherson_intrinsic_extrinsic_2023}, who, after extensive analysis of the literature on the intrinsic-extrinsic distinction, conclude that ``Unfortunately, when we look more closely at the intuitive distinction, we find reason to suspect that it conflates a few related distinctions, and that each of these distinctions is somewhat resistant to analysis." We consider several of these related distinctions and conclude that they fail to unambiguously yield the result that RDMs are intrinsic. The intrinsic-extrinsic distinction as it stands is thus not suited for making fine-grained distinctions within quantum theory.

In section \ref{mechanisms}, we offer some positive arguments in favor of treating changes in the relevant global states to be real changes in subsystems like qubit $B$. Our arguments emphasize the essential role these states play in explaining facts about measurement outcomes. We therefore define a sufficient condition for a theory to exhibit nonlocal action: \textit{if a physical theory makes essential explanatory use of a global state of spacelike separated regions, then an action that changes that state is a nonlocal action, for that physical theory}. We then consider a number of physical theories, to determine whether they meet the condition or not. We argue that our condition gives intuitive results for all these theories. 

Finally, in section \ref{concil}, we offer two conciliatory notes to EQM. The first is that by avoiding actions that remotely change RDMs, EQM acquires some explanatory power. These benefits should not be described in terms of EQM avoiding nonlocal action at a distance. But they can be described in alternative terms. The second is that there are Everett-inspired many worlds theories that \textit{are} fully local, although they do away with universal wavefunctions and especially nonseparability.

\section{The instrincality argument}\label{argument}

\subsection{The basic argument}

A common objection to our line of thinking, rejects the idea that Alice's action makes a \textit{real change} to Bob's system (qubit $B$). According to the objection, Alice's action does not make a real change to $B$, because she does not change any \textit{intrinsic} properties of $B$. She instead only makes \textit{extrinsic} or \textit{relational} changes to $B$, which are not really changes to $B$ itself. However, the intrinsic-extrinsic distinction is rarely spelled out in this argument. What is typically offered instead are classical analogies, which we argue are not compelling. 

One of the earliest versions of this argument can be found in \textcite{Timpson2002}, which is worth quoting at length:

\begin{quote}
``It seems that this appearance of non-locality is again not genuine, however. What have changed as a result of Alice’s measurement are the relative states of Bob’s system; that is, roughly, relational properties of his system. It is no mystery that relational properties can be affected unilaterally by operations on one of the relata and it certainly does not connote non-locality. (Compare ‘x is heavier than y’; we might make y heavier by adding weights, so that this statement becomes false, but this would not indicate a non-local effect on x.) [...] That is, the
genuine change is in fact all on Alice’s side. The upshot of this discursion is that we are indeed justified in saying that there is no non-locality in Everett. Although local unitary operations can have remarkable effects on the global state in the presence of entanglement, these effects are not such as to impute non-locality. There are no effects on local and non-relational properties of separated systems and, we have suggested, effects on relational properties [...] do not imply non-locality."
\end{quote}

We agree with the classical analogy in the following sense: rendering `x is heavier than y' false, by increasing y's weight, need not be a real change to x, so need not entail nonlocality. But it is quite a leap to go from this to the claim that the same holds for changes in global entangled states in quantum mechanics. What's needed is a clear account of this ``relational/non-relational" (or ``extrinsic/intrinsic") distinction. The account should not only group together the classical weight example and global entanglement states under the category `relational' in a natural way, but it also must ensure that the right things fall under the `non-relational' or `intrinsic' category. In particular, since changes to the RDMs of remote systems count as nonlocal changes, it must be that RDMs unambiguously fall under this category. 

In the later work of Timpson and Wallace, RDMs being intrinsic becomes a basic principle of what they call ``spacetime state realism". As \textcite[p299]{wallace2012emergent} writes:

\begin{quote}
    ``Timpson's and my proposal [...] is very straightforward: just take the density operator [RDM] of each subsystem to represent the intrinsic properties which that subsystem instantiates, just as the field values assigned to each spacetime point in electromagnetism represented the (electromagnetic) intrinsic properties which that point instantiated."
\end{quote}

However, no definition of `intrinsic' is offered, there is only this analogy with electromagnetism (see also \textcite[p709]{wallace2010quantum}).

The intrinsicality argument assumes that in EQM, actions only change relational/extrinsic properties of remote systems, and never intrinsic properties of remote systems. It concludes that EQM is local, in contrast to collapse theories. Before evaluating the  argument, an important terminological point is in order. \textcite[p293]{wallace2012emergent} defines \textit{action at a distance} to be what occurs when “given two systems $A$ and $B$
which are separated in space, a disturbance to A causes an immediate
change in the state of $B$, without any intervening dynamical
process connecting $A$ and $B$”. Furthermore, Wallace defines \textit{nonseparability}, to be what occurs when, ``given two regions A and B, a complete specification of the states of A and B separately fails to fix the state of the combined system, in addition to the facts about the two individual systems''. Wallace argues that while EQM exhibits nonspearability, it fails to exhibit action at a distance. 

Crucially, Wallace refers to nonseparability and action at a distance as two types of ``nonlocality". However, he also says that ``Nonseparability is a matter, not of dynamics, but of ontology" and concludes that EQM is ``a theory of local interactions and nonlocal states". Similarly, Timpson (this volume) distinguishes ``kinematic nonlocality,'' from ``dynamic nonlocality'' such that EQM is said to exhibit only the former. In these terms, our argument can be understood as concluding that EQM exhibits dynamical nonlocality, in certain situations, as a result of kinematic nonlocality.\footnote{\textcite{Vaidman2016} makes a related distinction between \textit{nonlocal action at a distance} and \textit{nonlocal objects}, and argues that EQM only has the latter. The former requires changes in `local descriptions' (instead of `intrinsic properties') by remote actions. Local descriptions are then identified with RDMs. We see no reason why actions on subsystems that change \textit{global descriptions} shouldn't also count as action at a distance.} 

The key question here is what counts as ``the state of B" in Wallace's definition of action at a distance. Defenders of the intrinsicality argument apparently want to restrict these states to ``intrinsic" properties, which in EQM include RDMs. But from our perspective, if $A$ and $B$ exhibit nonseparability, and a disturbance by Alice on $A$ immediately changes the nonseparable state of $A$ and $B$, then why is this not an immediate ``change in the state of $B$"? Let us then delve deeper into this issue.

\subsection{The Socrates-Xanthippe analogy}

One of the clearest and most detailed expositions of the intrinsicality argument can be found in a recent paper by \textcite{ney2023argument}. Here, Ney improves upon previous formulations, especially by distinguishing different metaphysical interpretations of EQM, and showing how the intrinsicality argument applies to each one. However, here too no explicit definition of `intrinsic' or `extrinsic' are offered, the distinction is taken for granted and RDMs are assumed to count as intrinsic properties. As with \textcite{Timpson2002}, what is offered instead is a classical analogy. In Ney's example, we consider the immediate effects of Socrates' death, on his wife Xanthippe, who is far away.\footnote{See also Ney (this volume) for a related example.} Xanthippe is immediately widowed. But this is not a real change to Xanthippe because it does not change her intrinsic properties. It therefore cannot be deemed action at a distance. 

\textcite{faglia2024non}, who also defends the intrinsicality argument, follows Ney, by offering the same analogy, expressed as follows:

\begin{quote}
``When Socrates drinks the hemlock and dies, Xanthippe instantly becomes a widow. Plausibly, Socrates drinking the hemlock causes Xanthippe to become a widow. Although Socrates and Xanthippe may be spacelike separated, the scenario evidently involves no violation of Local Causality because the change in Xanthippe only involves extrinsic properties, namely the extrinsic property of \textit{being a widow}, and thus it
is not localized in Xanthippe’s region."
\end{quote}

The idea is that we should think the same thing about Alice, and the changes she makes to $B$. By changing the global quantum state, Alice only changes extrinsic properties of $B$. Nonlocal action requires changes in intrinsic properties, but that would require changes to $B$'s RDM. 

We wish to raise two types of challenges to this argument. The first is that the intrinsic-extrinsic distinction is not sufficiently well-defined to reliably determine whether a physical theory exhibits nonlocal action. The second is that insofar as we have even a rough idea of that distinction, it is far from clear why RDMs count as intrinsic properties. We argue these two points in the remainder of section \ref{argument}. In section \ref{concept} we provide some background to the distinction to show how widely contested it is in the philosophical literature, even for its application in classical metaphysics. Then in section \ref{relational}, we consider whether `intrinsic' can be simply defined as `non-relational', and raise several doubts about that. Here we emphasize the difficulty in obtaining the result that RDMs are intrinsic. Finally in section \ref{other} we consider two other definitions of `intrinsic' that have been proposed in the literature, and show that they do not help the intrinsicality argument for locality in EQM.

\subsection{The intrinsicality concept}\label{concept}

We begin with some of the history behind the intrinsic-extrinsic property distinction. We do this partly to emphasize the fact that the distinction has been developed primarily in the context of philosophical applications that bear little relevance to quantum mechanics. There is therefore little reason to expect that it must apply cleanly, and in an informative way, to quantum properties.\footnote{Our point here is similar to one made by \textcite[p89]{maudlin2019philosophy}, who, in his discussion of the reality of the quantum state, laments the project of trying to fit the quantum state into one or another \textit{Aristotelian} metaphysical category.}  

Perhaps one of the earliest explicit uses of the distinction in philosophy is in value theory. In particular, \textcite[Sec. 18]{moore1903principia} was one of the first philosophers to explicitly put the term `intrinsic' to use. He distinguished things that are \textit{intrinsically good} i.e. good in themselves, from things that are good as a causal means to other things.\footnote{This corresponds to a common gloss on the more general notion of intrinsic property: ``A thing has its intrinsic properties in virtue of the way that thing itself, and nothing else, is. Not so for extrinsic properties"(\textcite[p197]{Lewis1983}).} This version of the intrinsic-extrinsic distinction has been challenged, on the grounds that it is not sufficiently well-defined (\textcite{Feldman1998}, \textcite{Kagan1998}), and on the grounds that insofar as it can be roughly defined, nothing much ends up being intrinsic (\textcite{Krebs1999}). These challenges mirror the challenges that we will raise for the general intrinsic-extrinsic distinction. In particular, whether it is well-defined, and whether anything - especially RDMs - count as intrinsic. 

The distinction would later come to underpin discussions of \textit{real change} versus merely ``Cambridge" change (\textcite{Geach1969}) and has been invoked in metaphysical debates about persistence, supervenience, and the recombination of possible states (\textcite{Lewis1983, Jackson1998, Kim1982}). These applications often concern notions of value, modality, and identity over time in classical contexts—domains that differ fundamentally from the mathematical structures of quantum mechanics. Consequently, there is little reason to expect that the intrinsic-extrinsic distinction, as developed in these contexts, should map cleanly onto quantum notions like RDMs and global entangled states.

There are typically two ways in which the distinction is introduced in the literature (\cite{marshall_weatherson_intrinsic_extrinsic_2023}). One is by giving \textit{paradigm examples} of intrinsic and extrinsic properties. Paradigm extrinsic properties are apparently easy to come by. Being a wife or being widowed, weighing more than Socrates, or being one mile from the town center, all seem like mere relations to other things, and therefore not intrinsic properties. But uncontroversial intrinsic properties are much harder to come by. Lewis (\citeyear{Lewis1983, Lewis1983Universals, Lewis1988}), for example, has insisted that shape properties are intrinsic, but others have held that an object’s shape depends on the curvature of the space in which it is located, and this might not even be intrinsic to that space (\textcite{Nerlich1979}), let alone the object (\textcite{Bricker1993, McDaniel2007, Skow2007}). Similar considerations have been raised for an object's solidity (\textcite{mcqueen2016tests}) and mass (Dasgupta \citeyear{dasgupta2013absolutism, dasgupta2021relationalist}). It is difficult to find any noncontroversial intrinsic property in the literature. This makes it difficult to establish that RDMs are intrinsic properties by comparing them to `paradigm' intrinsic properties.

The other way of introducing the distinction is by trying to define it, at least roughly. Unfortunately, the intrinsicality literature offers a bewildering variety of definitions. We will consider some options in what follows, and raise problems for their potential use in the argument for EQM's locality.

\subsection{Intrinsic as non-relational}\label{relational}

A common way to characterize intrinsic properties is by contrast with \textit{relational} properties. On this view, an intrinsic property of an object is one whose possession does not consist in standing in any relation to any other objects. However, this approach faces counterexamples. Many properties are both relational and plausibly intrinsic. For instance, consider your property of being a cordate (a creature with a heart). This would seem to be a property you have intrinsically, in virtue of the way you are. Yet it is also a relation, specifically, a part-whole relation: you have a proper part that is a heart. Since you are not identical to your heart, this is a relation between distinct objects. \textcite{marshall_weatherson_intrinsic_extrinsic_2023} offer a related counterexample: most humans have the property of having longer legs than arms, a fact that depends on internal relations between their limbs but does not seem extrinsic in any relevant sense. 

Different options have been suggested for the non-relational account of intrinsicality. For example, \textcite{Humberstone1996} considers defining a relational property such that if an object has the property, then it bears some relation to \textit{a non-part of it}. But consider the property of \textit{not} being within a mile of a gas station. This is relational, but does not consist in bearing a relation to any non-part, since one has it in a world without any gas stations. 

\textcite{marshall_weatherson_intrinsic_extrinsic_2023} consider another strategy, which says that a relational property can only relate \textit{wholly distinct things}, and that non-intrinsic properties are relational in this sense. But this won't help the argument for locality in EQM, since it will fail to render the relevant global states as extrinsic. The relevant global states involve relations between entities that are \textit{not} wholly distinct. In particular, entangled particles are not wholly distinct in that they are nonseparable: a complete specification of the states of $A$ and $B$ separately fails to fix the state of the combined system.

\textcite[fn 2, p. 44]{faglia2024non} may be the first attempt to defend the intrinsicality argument for locality in EQM with an explicit definition of `intrinsic', one which avoids some of these problems. Intrinsic properties are defined to be those properties “such that having them does not consist in being related, or failing to be related, in any way to any external object or objects”. The appeal to ``failing to be related" removes the gas station counterexample. The appeal to ``external objects" might seem to get around the problems raised by entangled objects that are not wholly distinct from each other. 

However, Faglia's definition raises its own problems. In particular, it is hard to see why RDMs \textit{must} count as intrinsic. Does $B$'s having the RDM that it has, \textit{consist in} being related to any external object, such as qubit $A$? It is left vague how we are to go about answering this question. One way to answer it would be to say: it depends on whether \textit{the calculation} of $B$'s RDM depends on its relation to $A$. The thought is, if $B$'s RDM is intrinsic to $B$, then I should not need to look at anything other than $B$ in order to figure out what that RDM is. But then, since one calculates $B$'s RDM by tracing $A$ out of the nonseparable state of $A$ and $B$, it follows that $B$'s RDM is an \textit{extrinsic} property of $B$. To calculate the RDM of \( B \), we start with the global density matrix \( \rho_{AB} = \ket{\Psi} \bra{\Psi} \), where \( \ket{\Psi} \) is the Bell state from equation (1). We then compute the partial trace over \( A \) in the computational basis:
\begin{equation}
\rho_B = \text{Tr}_A (\rho_{AB}) = \bra{0}_A \rho_{AB} \ket{0}_A + \bra{1}_A \rho_{AB} \ket{1}_A.
\end{equation}
Carrying out this operation explicitly, we find equation (2) from above, representing $B$'s RDM. Faglia's definition therefore fails to unambiguously render RDMs intrinsic. 

Here is a very different way of using Faglia's definition to conclude that RDMs are extrinsic. $B$'s RDM encodes \textit{probabilities}. To a first approximation, these are probabilities \textit{for an external system} that interacts with $B$ to obtain certain outcomes. However, this raises a deeper problem for the account, given that there is no generally accepted interpretation of probability in EQM. For a collapse theory, the probabilities encoded in an RDM could be understood as \textit{dispositions} or \textit{propensities} for the system to spontaneously collapse to a certain state (\textcite{frigg2007probability, lorenzetti2021refined}). Perhaps those dispositions could be interpreted as intrinsic properties that get carried around by the system. But there is no disposition to collapse in EQM. So then what are the probabilities \textit{probabilities of}? 

On some views, the answer might seem to render RDMs extrinsic. For example, according to \textcite{greaves2004understanding}, the Born rule is understood to be an agent's `caring measure', since the measure quantifies the extent to which (for decision-making purposes) the agent cares about what happens on any given branch. One may then conclude that since $B$'s having the RDM it has partly consists in the care an external agent has for branches associated with $B$, the RDM must be an extrinsic property of $B$. One could make similar points if one thinks of quantum probabilities in terms of how a rational agent should bet (\textcite{deutsch1999quantum}, \textcite{wallace2012emergent}), or in terms of an agent's self-location uncertainty (\textcite{mcqueen2019defence}, \textcite{sebens2018self}). Other interpretations of the Born rule in EQM, however, may make RDMs seem intrinsic, such as \textcite{vaidman1998schizophrenic}'s measure of existence. Our point is not that RDMs should be understood as extrinsic in EQM, but that given the definitions offered so far, there is simply no clear fact of the matter - one could go either way - and the probability problem in EQM exacerbates this.

\subsection{Grounding and contractionist definitions}\label{other}

We conclude with brief mention of two other influential definitions of `intrinsic'. First, there have been several attempts to define intrinsicality in terms of \textit{metaphysical grounding}, including \textcite{witmer2005intrinsicality}, \textcite{trogdon2009monism}, \textcite{rosen2010metaphysical}, \textcite{bader2013towards}, and \textcite{witmer2014simple}. Here is Rosen's definition:

\begin{quote}
\textit{F} is an intrinsic property iff, necessarily, for any \( x \): (i) if the ascription of \( F \) to \( x \) is grounded by a fact with constituent \( y \), then \( y \) is part of \( x \); (ii) if the negation of the ascription of \( F \) to \( x \) is grounded by a fact with constituent \( y \), then \( y \) is part of \( x \).
\end{quote}

To apply this to our example: if $B$'s RDM is an intrinsic property of $B$, then it \textit{cannot} be that the ascription of that RDM to $B$ is grounded by a fact with constituent $y$, where $y$ is not a part of $B$. But arguably, the ascription of the RDM to $B$ \textit{is} grounded by a fact with constituent $A$. The fact is the global entangled state of $A$ and $B$. And this is the more fundamental fact upon which the fact about $B$'s RDM is grounded. After all, one starts with the global state, and then derives $B$'s RDM from that, by tracing out $A$ (equation (5)).

Second, there have been attempts to define intrinsicality in terms of \textit{contraction}. An early version of the idea was suggested by \textcite[p197]{Lewis1983}: ``If something has an intrinsic property, then so does any perfect duplicate of that thing; whereas duplicates situated in different surroundings will differ in their extrinsic properties." There have been various attempts to make the idea more precise. The contractionist version of this idea situates duplicates in empty worlds, to see what properties they retain. Roughly, a thing’s intrinsic properties are those properties it would have if it were alone in the world. This is an idea developed by \textcite{vallentyne1997intrinsic}, who defines a contraction of a world as “a world ‘obtainable’ from the original one solely by ‘removing’ objects from it.” A property is then intrinsic to an object if and only if removing the rest of the world doesn’t change whether the object has the property. More formally, Vallentyne defines an x-t contraction as ‘removing’ all objects wholly distinct from x, all spatial locations not occupied by x, and all times (temporal states of the world) except t, from the world. Vallentyne then claims that 

\begin{quote}
$F$ is an intrinsic property iff, for any thing $x$, for any time $t$, and for any worlds $u$ and $w$ such that $u$ is an $x$-$t$ contraction of $w$, $x$ has $F$ at $t$ at $u$ iff $x$ has $F$ at $t$ at $w$.  
\end{quote}

To apply this to our example, $B$'s RDM is intrinsic if $B$ retains that RDM when it is made lonely (i.e. by ``contracting the universe" by hypothetically deleting everything else in existence, including qubit $A$). But what sense can be made of making $B$ lonely when it is entangled with $A$? Recall that we calculate $B$'s RDM by tracing out $A$. If we were to hypothetically `delete' $A$, there is no reason to suppose that $B$ would survive given their entanglement. One could insist on this, but it is unclear what makes it so. In particular, there is no reason to suppose that what remains would be $B$ with the same RDM as the one that we calculated using $A$ - why not the state of $B$ just before it became entangled with $A$?

To conclude, we are not trying to insist that these definitions unambiguously render RDMs extrinsic. Our point is simply that one could equally well go either way. The intrinsic-extrinsic distinction was developed in a non-quantum context, such that there is no guarantee that it can be straightforwardly extended to the quantum context. Of course, it may be possible to define a more sophisticated definition of `intrinsic' that does the job. We leave this, then, as an open challenge for defenders of the intrinsicality argument.

\section{Global state changes as real changes}\label{mechanisms}

\subsection{The basic idea}

In this section, we try to make a positive case for treating the change in the global state induced by Alice as a real change to qubit $B$, and therefore, as a case of nonlocal action at a distance. To do this, we try to sketch a plausible \textit{sufficient condition}, for nonlocal action. 

We assume that whether or not a theory is local depends on the theory's ontology (\textcite[Sec. 4]{waegell2023generative}).
This contrasts with operational conceptions of locality, for example in terms of no-signaling. For an action that changes a global state to count as a nonlocal action, the global state must be an \textit{essential explanatory mechanism} of the theory. In EQM, the global state is needed to explain why, in an anti-correlated Bell state, Alice’s measuring spin-up ensures that her future self encounters only the branch where Bob measures spin-down. 

Compare this with the change in Xanthippe, when she becomes a widow. We agree that the scenario does not involve nonlocal action at a distance. But we think that has little to do with the fact that the change in Xanthippe is a change in ``extrinsic" properties. Rather, it has to do with the fact that the change in Xanthippe is not a change in an explanatorily essential state, according to any reasonable physical theory.

To see this, note that we need not appeal to any physical ``widowing" relation to explain anything that we observe. Global states in EQM could not be more different than this. For in EQM we have global state changes, such as $B$ becoming entangled with Alice, when Alice performs her measurement (at spacelike separation from $B$). The global state changes at the instant Alice measures (regardless of which Lorentz frame we consider). It is the ineliminable explanatory role that such global states play in EQM that makes them relevant to whether or not EQM is local or nonlocal - they cannot simply be swept under the rug due to being ``extrinsic''.

\subsection{The Socrates-Xanthippe analogy revisited}

\begin{figure}[h!]
    \centering
    \includegraphics[width=3.7in]{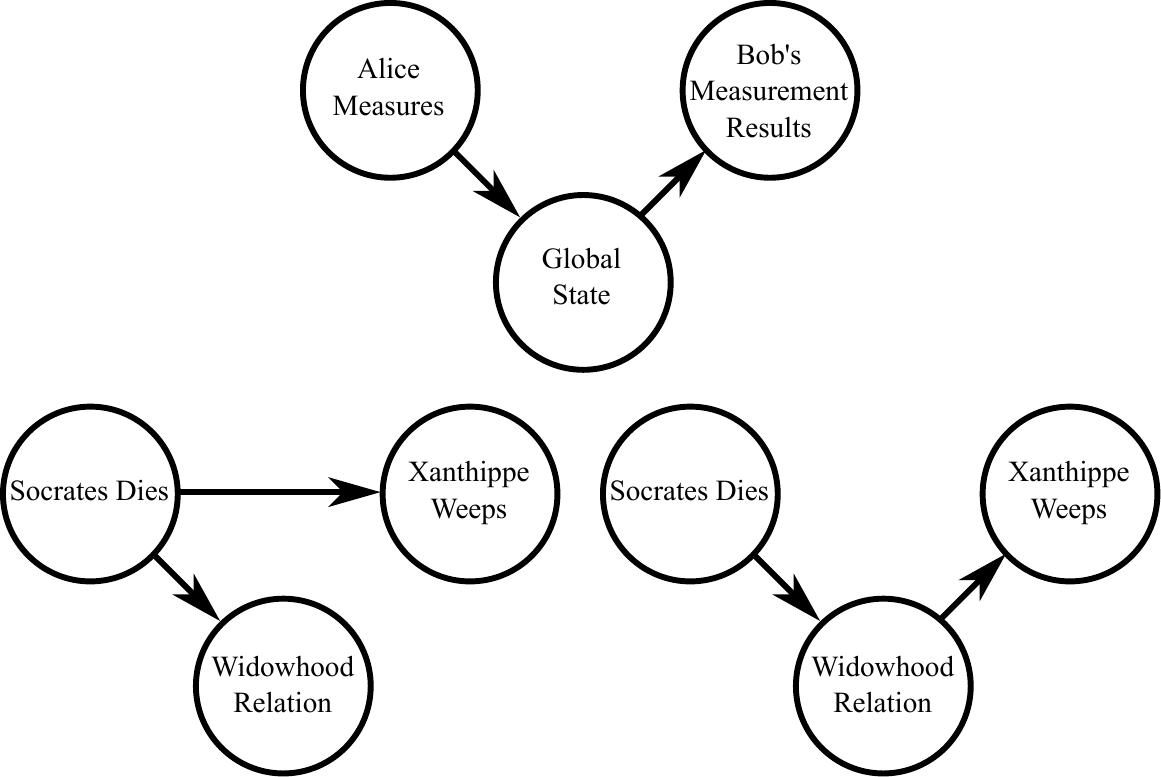}
    \caption{\small Causality Diagrams. Top: Alice's action affects the global state, (which includes qubit $B$), which is essential to explaining Bob's measurement outcomes. Right: The unfamiliar strange theory. Socrates' execution affects the widowhood relation,  which is essential to explaining why Xanthippe weeps. Left: The familiar classical case. The widowhood relation does not explain why Xanthippe weeps, she weeps because she hears of Socrates' death.}
    \label{DAGs}
\end{figure}

We can further illuminate the basic idea by considering a strange physical theory where becoming a widow \textit{is} an integral part of the theory's explanatory machinery. In particular, we can imagine that the property of being a widow plays an essential role in explaining some observations, such that we cannot eliminate it by instead appealing to subluminal signals sent from the execution event. In the strange theory, we imagine that when Xanthippe weeps upon seeing Socrates' deceased body, her weeping is not caused by signals from the body, it is caused by the delocalized widowhood relation. In that case, instantaneously becoming a widow \textit{should} suggest nonlocality, independently of whether or not we consider it to be a change in extrinsic or intrinsic properties. See Figure \ref{DAGs}.

To see why this should suggest nonlocality, we need to look at the details of the widowhood relation.  This relation cannot be defined without reference to both Socrates and Xanthippe, so it is not localized in either of their regions.  If this delocalized relation plays an essential role in the physical explanation of observations, then the physical theory is nonlocal. An entangled global state also cannot be defined without reference to all of the entangled subsystems, so this is another relation that cannot be localized into the regions of the individual systems, and so theories with such global states are nonlocal for the same reason. 

The intuition that entanglement relations are consistent with locality could perhaps be traced to a conception of them, which treats them as ontologically distinct from RDMs. This is suggested by an analogy provided by \textcite[p304]{wallace2012emergent}, where ``picturesquely, we can think of entanglement between [subsystem] states [represented by their RDMs] as a string connecting those states, representing the nonlocal relation between them". But any classical string could be cut and separated from the objects at its ends, and then both objects, and the cut string, would continue to exist, and have their own well-defined properties.  The RDMs cannot be detached from the entanglement relation in this way, which is why the classical analogy is misleading.  The RDMs of Bob and Alice's particles are fully defined by the entanglement relation and fully constituted by it. For the entanglement relation just \textit{is} the Bell state. The mere fact that RDMs describe the local measurement statistics of individual systems does not mean they exist independently from the entanglement relations. If an entanglement relation instantaneously vanished, then all the reduced density matrices of all the entangled systems it describes, no matter how far apart, would also vanish simultaneously.  In contrast, there is nothing like this in classical physics, where all extended objects can be separated into their constituent parts, and those parts have their own independent properties.

\subsection{Application to various physical theories}

In this section, we consider whether a number of different physical theories satisfy our condition or not. We argue that our condition yields intuitive verdicts. In the process, we consider how one might try to object to our account by identifying counterexamples i.e., local theories with explanatorily essential global states. We will argue that the theories considered do not offer any clear counterexamples. 

We begin with a possible counterexample. One might think that global states which specify \textit{distance relations} are explanatory physical relations in most physical theories, whether or not those theories are local or nonlocal. For example, the distance between two masses is a global fact that can be changed by moving just one of the masses. This need not be nonlocal action at a distance, even though it might still be useful in explanation. Our response is that in local theories, distance relations are not explanatorily \textit{essential}. Meanwhile, in theories that do satisfy our condition, like Newtonian gravity, distance relations play an essential role in the laws ($r^2$). 

Our sufficient condition for nonlocal action is not met by classical mechanics with special relativity but without gravity. This is because this theory explains everything using local separable states.  The theory uses separable states because the time-evolution laws at a given location can depend only on the contents of that location's past light cone. The theory therefore requires that the light cone's contents be separable from what's outside of that light cone. And everything inside that light cone must be separable too, since the same argument applies to locations inside that light cone. Thus, there are no global states of spacelike separated regions with essential explanatory power in this theory.

Likewise in classical electrodynamics, all influences are mediated by influences that propagate at or below $c$, and the instantaneous distance $r$ between remote objects never plays any essential explanatory role.  This is because the retarded electromagnetic potentials at a given event $(\vec{x},t)$, obtained in the properly relativistic Lorenz gauge, depend not on the instantaneous location of the source object, but on where it was when it emitted a signal which propagates at $c$ along a geodesic and arrives at $(\vec{x},t)$.  In quantum electrodynamics, these signals are characterized as `virtual photons.' Influences in general relativity appear to work this way also, but the theory may be nonlocal for other reasons, a point we return to in a moment. 

Our sufficient condition for nonlocal action is clearly satisfied by Newton's theory of gravity, because the gravitational force in this theory depends on a global state, which thus plays an ineliminable role in how the theory explains what happens.  In this case, the global state in question is the instantaneous distance $r$ between two masses, which is well-defined in nonrelativistic classical physics with absolute time.  When either mass moves, the global distance relation changes, and because this global state plays an essential role in how this physical theory determines what happens next, our condition for nonlocality is met.

Two theories that straightforwardly satisfy our condition are Bohmian mechanics and collapse theories. Bohmian mechanics satisfies our condition for nonlocality, because it uses a universal wavefunction as a global state, which is changed by Alice's measurement, and which is essential to explain the trajectory of the Bohmian hidden variable through configuration space. Collapse theories satisfy our condition because of the role of delocalized entangled states in explaining collapse events. 

Finally, we consider general relativity (GR). At first, it might seem that our condition renders GR nonlocal. Like EQM, GR is nonseparable: if we accept the usual view of diffeomorphism invariance (\textcite{Adlam2024Relational}, \textcite{Wallace2002TimeDependent}, \textcite{Earman2002McTaggart}), only the relational quantities or integrals over all of spacetime are part of the ontology. But unlike EQM, it's hard to identify anything in the theory that you could consider as a subsystem state (like an RDM) that would be changed by an action (\textcite{rovelli2023philosophical}, \textcite{Rickles2008BackgroundIndependence}). For our condition to be met, we also need well-defined actions, which change global states. But it remains controversial how actions can be modeled in GR.

\section{Conclusion}\label{concil}

In this paper, we have raised a challenge to the view that EQM avoids nonlocal action at a distance. EQM allows actions on a system that change the global state of the system and its remote entangled partners. We criticised the idea that this is not a real ``intrinsic" change to the system. We then tried to clarify when actions that change global states count as nonlocal actions: when the global states are essential explanatory mechanisms of the theory. In EQM, the global state is needed to explain why, in an anti-correlated Bell state, Alice's and Bob's successors see results that are properly correlated. 

However, we do not think our considerations entirely undercut the argument for EQM based on its purported locality, for two reasons. First, by removing the special type of nonlocality that comes with the collapse of the wave function, that is, the action at a distance that changes remote RDMs, EQM offers a much better account of otherwise paradoxical quantum experiments (\textcite{mcqueen2020many}). Second, although we concluded that nonseparable theories such as EQM are nonlocal, it does not follow that all many worlds theories are nonlocal. For it has been shown that many worlds theories need not involve nonseparable global states at all.\footnote{\textcite{deutsch2000information, deutsch2012vindication, brassard2013can,  waegell2017locally, waegell2018ontology, brassard2019parallel, bedard2021cost,  bedard2021abc, kuypers2021everettian, waegell2023local, bedard2023teleportation, waegell2024toward,  bedard2024local}.} Many worlds theories can therefore still enable one to avoid nonlocality, and Bell's theorem still needs to be supplemented explicitly with a ``one world" assumption to capture it (\cite{WAEGELL2020}).

\section*{Acknowledgments}
We would like to thank Emily Adlam, Alyssa Ney, and an anonymous referee, for helpful feedback.  This project/publication was made possible through the support of Grant 63209 from the John Templeton Foundation. The opinions expressed in this publication are those of the authors and do not necessarily reflect the views of the John Templeton Foundation.

\printbibliography

\end{document}